\documentclass[a4paper,10pt,oneside]{scrartcl}
\usepackage{graphicx}
\usepackage{epsfig}
\usepackage{verbatim}
\usepackage{amsmath, amssymb, amsthm, graphics}

\usepackage[inner=3.4cm,outer=2.5cm,top=4cm,bottom=3.2cm]{geometry}
\usepackage[T1]{fontenc}  
\usepackage{mathptmx}  
\usepackage{caption}
\newcommand{\rubric}[1]{%
\bigskip
\textbf{\large #1}  

\medskip}

\theoremstyle{plain}

\theoremstyle{definition}

\theoremstyle{remark}

\usepackage{subfigure}

\newcommand{\D}{{\mathrm{d}}}

\newcommand{\bfx}{ \mathbf{x}}
\newcommand{\bfy}{ \mathbf{y}}
\newcommand{\bfv}{ \mathbf{v}}
\newcommand{\bfu}{ \mathbf{u}}
\newcommand{\bff}{ \mathbf{f}}
\newcommand{\bffeq}{\bff^{\mathrm{eq}}}

\begin{document}

\textbf{\Large Add-ons for Lattice Boltzmann Methods:
Regularization, Filtering and Limiters}

\bigskip
\textbf{R.A.~Brownlee, J.~Levesley, D.~Packwood and A.N.~Gorban
}

\bigskip
\emph{Department of Mathematics, University of Leicester, Leicester LE1 7RH, UK}

\bigskip~
\hspace*{1cm}\begin{minipage}{12.75cm}{Abstract: We describe how
regularization of lattice Boltzmann methods can be achieved by
modifying dissipation. Classes of techniques used to try to improve
regularization of LBMs include flux limiters, enforcing the exact
correct production of entropy and manipulating non-hydrodynamic
modes of the system in relaxation. Each of these techniques
corresponds to an additional modification of dissipation compared
with the standard LBGK model. Using some standard 1D and 2D
benchmarks including the shock tube and lid driven cavity, we
explore the effectiveness of these classes of methods.
}\end{minipage}

\vspace*{2\baselineskip}
\textbf{\large Keywords:} Lattice Boltzmann,
Dissipation,
Stability,
Entropy,
Filtering,
Multiple Relaxation Times (MRT),
Shock Tube,
Lid-Driven Cavity

\vspace*{2\baselineskip}

\textbf{\large 1. INTRODUCTION}
\setcounter{section}{1}
\setcounter{subsection}{0}
\setcounter{subsubsection}{1}

Lattice Boltzmann methods (LBM) are a class of discrete
computational schemes which can be used to simulate
hydrodynamics and more\cite{Succi}. They have been proposed as
a discretization of Boltzmann's kinetic equation. Instead of
fields of moments $M$, the LBM operates with fields of discrete
distributions $f$.

All computational methods for continuum dynamics meet some
troubles with stability when the gradients of the flows become
too sharp. In Computational Fluid Dynamics (CFD) such
situations occur when the Mach number $Ma$ is not small, or the
Reynolds number $Re$ is too large. The possibility for using grid
refinement is bounded by computational time and memory
restrictions. Moreover, for nonlinear systems with shocks
then grid refinement does not guarantee  convergence to the
proper solution. Methods of choice to remedy this are based on the {\em
modification of dissipation} with limiters, additional
viscosity, and so on \cite{Wess}. All these approaches combine
high order methods in relatively quiet regions with
low order regularized methods in the regions with large
gradients. The areas of the high-slope flows are assumed to be
small but the loss of the order of accuracy in a small region
may affect the accuracy in the whole domain because of the
phenomenon of error propagation. Nevertheless, this loss of
accuracy for systems with high gradients seems to be
unavoidable.

It is impossible to successfully struggle with some spurious effects without a local decrease in the order of accuracy. In a more formal
setting, this has been proven. In 1959, Godunov \cite{Godunov} proved
that a (linear) scheme for a partial differintial equation
could not, at the same time, be monotone and second-order
accurate. Hence, we should choose between spurious oscillations
in high order non-monotone schemes and additional dissipation
in first-order schemes. Lax \cite{Lax} demonstrated that
un-physical dispersive oscillations in areas with high slopes
are unavoidable due to discretization. For hydrodynamic
simulations using the standard LBGK model (Sec. \ref{sec:bac})
such oscillations become prevalent especially at high $Re$ and
non-small $Ma$. Levermore and Liu used differential
approximation to produce the ``modulation equation" for the
dispersive oscillation in the simple initial-value Hopf problem
\cite{LeverLiu1996} and demonstrated directly how for a
nonlinear problem a solution of the discretized equation does
not converge to the solution of the continuous model with high
slope when the step $h \to 0$.

Some authors expressed a hope that precisely keeping the
entropy balance can make the computation more ``physical", and that
thermodynamics can help to suppress nonphysical effects. Tadmor
and Zhong constructed a new family of entropy stable difference
schemes which retain the precise entropy decay of the
Navier--Stokes equations and demonstrated that this precise
keeping of the entropy balance does not help to avoid the
nonphysical dispersive oscillations \cite{Tadmor2006}.

To prevent nonphysical oscillations, most upwind schemes employ
limiters that reduce the spatial accuracy to first order
through shock waves. A mixed-order scheme may be defined as a
numerical method where the formal order of the truncation error
varies either spatially, for example, at a shock wave, or for
different terms in the governing equations, for example,
third-order convection with second-order diffusion
\cite{Roy2003}.

Several techniques have been proposed to help suppress these
pollutive oscillations in LBM, the three which we deal with in
this work are entropic lattice Boltzman (ELBM), entropic
limiters and generalized lattice Boltzmann, also known as
multiple relaxation time lattice Boltzmann (MRT). Where
effective each of these techniques corresponds to an additional
degree of complexity in the dissipation to the system, above
that which exists in the LBGK model.

The Entropic lattice Boltzmann method (ELBM) was invented first
in 1998 as a tool for the construction of single relaxation
time LBM which respect the  $H$-theorem \cite{Karlin2}. For
this purpose, instead of the mirror image with a local
equilibrium as the reflection center, the entropic involution
was proposed, which preserves the entropy value. Later, it was
called the {\it Karlin-Succi involution} \cite{Gorban}. In
2000, it was reported that exact implementation of the
Karlin-Succi involution (which keeps the entropy balance)
significantly regularizes the post-shock dispersive
oscillations \cite{Ansumali}. This regularization seems very
surprising, because the entropic lattice BGK (ELBGK) model
gives a second-order approximation to the Navier--Stokes
equation similarly to the LBGK model (different proofs of that
degree of approximation were given in \cite{Succi} and
\cite{Brownlee}).

Entropic limiters \cite{Brownlee} are an example of flux
limiter schemes \cite{Brownlee,Kuzmin,Ricot}, which are
invented to combine high resolution schemes in areas with
smooth fields and first order schemes in areas with sharp
gradients. The idea of flux limiters can be illustrated by the
computation of the flux $F_{0,1}$ of the conserved quantity $u$
between a cell marked by 0 and one of two its neighbour cells
marked by  $\pm 1$:
\begin{equation}
   F_{0,1}=(1-\phi (r))f^{\rm low}_{0,1} + \phi (r) f^{\rm high}_{0,1},
\end{equation}
where $f^{\rm low}_{0,\,  1}$, $f^{\rm high}_{0,\,  1}$ are low
and high resolution scheme fluxes, respectively,
$r=(u_0-u_{-1})/(u_1-u_0)$, and $\phi (r)\geq 0$ is a flux
limiter function. For $r$ close to 1, the flux limiter function
$\phi (r)$ should be also close to 1. Many flux limiter schemes
have been invented during the last two decades~\cite{Wess}. No
particular limiter works well for all problems, and a choice is
usually made on a trial and error basis. Particular examples of
the limiters we use are introduced in Section
\ref{sec:filtering}.

MRT has been developed as a true generalization of the
collisions in the lattice Bhatnagar--Gross--Krook (LBGK) scheme
\cite{Dellar,Lallemand} from a one parameter diagonal relaxation
matrix, to a general linear operation with more free
parameters, the number of which is dependent on the particular
discrete velocity set used and the number of conserved
macroscopic variables. Different variants of MRT have been
shown to improve accuracy and stability, including in our
benchmark examples \cite{Luo} in comparison with the standard
LBGK systems.

The lattice Boltzmann paradigm is now mature, and explanations
for some of its successes are available. However, in its
applications it approaches the boundaries of the applicability
and need special additional tools to extend the area of
applications. It is well-understood that near to shocks, for
instance, special and specific attention must be paid to avoid
unphysical effects. In this paper then, we will discuss a
variety of {\em add-ons} for LBM and apply them to a variety of
standard 1D and 2D problems to test their effectiveness. In
particular we will describe a family of entropic filters and
show that we can use them to signficantly expand the effective
range of operation of the LBM.

\rubric{2. BACKGROUND} \setcounter{section}{2}
\setcounter{subsection}{0} \setcounter{subsubsection}{1}
\label{sec:bac} Lattice Boltzmann methods can be derived
independently by a discretization Boltzmann's equation for kinetic
transport or by naively creating a discrete scheme which matches
moments with the Maxwellian distribution up to some finite order.

In each case the final discrete algorithm consists of two
alternating steps, advection and collision, which are applied to $m$
single particle distribution functions $f_i\equiv f_i(\bfx,t), (i =
1\ldots m)$, each of which corresponds with a discrete velocity
vector $\bfv_i, (i=1\ldots m)$. The values $f_i$ are also sometimes
known as \emph{populations} or \emph{densities} as they can be
thought of as representative of the densities of particles moving in
the direction of the corresponding discrete velocities.

The advection operation is simply free flight for the discrete time
step $\epsilon$ in the direction of the corresponding velocity
vector,

\begin{equation}
f_i(\bfx,t+ \epsilon) = f_i(\bfx - \epsilon \bfv_i,t) .
\end{equation}

The collision operation is instantaneous and can be different for
each distribution function but depends on every distribution
function, this might be written,

\begin{equation}
f_i(\bfx) \rightarrow F_i( \{ f_i(\bfx) \} ).
\end{equation}

In order to have a slightly more compact notation we can write these
operations in vector form, in the below equation it should be
inferred that the $i$th distribution function is advecting along its
corresponding discrete velocity,
\begin{equation}
\bff(\bfx,t+ \epsilon) = \bff(\bfx - \epsilon \bfv_i,t) ,
\end{equation}
\begin{equation}
\bff(\bfx) \rightarrow F( \bff(\bfx) ).
\end{equation}

To transform our vector of microscopic variables at a point in space
$\bff(\bfx)$  to a vector of macroscopic variables $M(\bfx)$ we use
a vector of linear functions $M(\bfx) = m(\bff(\bfx))$. In the
athermal hydrodynamic systems we consider in this work the momemts
are density $\rho$ and momentum density $\rho \bfu$, $\{\rho, \rho
\bfu \}(\bfx) = \sum_i \{1, \bfv_i\} f_i(\bfx)$. These macroscopic
moments are conserved by the collision operation, $m(\bff) =
m(F(\bff))$.

The simplest and most common choice for the collision operation
$F$ is the
Bhatnagar-Gross-Krook(BGK)~\cite{Benzi,Chen,Higuera,Succi}
operator with over-relaxation
\begin{equation}
F(\bff) = \bff + \alpha\beta(\bffeq  - \bff).
\label{eq:CollisionIntegral}
\end{equation}
For the standard LBGK method $\alpha = 2 $ and $\beta \in [0, 1]$
(usually, $\beta \in [1/2, 1 ]$)  is the over-relaxation coefficient
used to control viscosity. For $\beta = 1/2$ the collision operator
returns the \emph{local equilibrium} $\bffeq$ and $\beta = 1$ (the
\emph{mirror reflection}) returns the collision for a liquid at the
zero viscosity limit. The definition of $\bffeq$ defines the
dynamics of the system, often it chosen as an approximation to the
continuous Maxwellian distribution. An equilibrium can also be
independently derived by constructing a discrete system which
matches moments of the Maxwellian up to some finite order. For
hydrodynamic systems often this finite order is chosen to be 2, as
this is sufficient to accurately replicate the Euler(non
dissipative) component of the Navier Stokes equations. For a
dissipative fluid with viscosity $\nu$ the parameter $\beta$ is
chosen by $\beta =\epsilon/(2\nu+ \epsilon)$.

Each of the techniques we test in this paper can be introduced as
developments of the generic LBGK system and such a presentation
follows in the next sections.

\rubric{3. ENTROPIC LATTICE BOLTZMANN} \setcounter{section}{3}
\setcounter{subsection}{0} \setcounter{subsubsection}{1}

\subsection{LBM with $H$ theorem}

In the continuous case the Maxwellian distribution maximizes
entropy, as measured by the Boltzmann $H$ function, and
therefore also has zero entropy production. In the context of
lattice Boltzmann methods a discrete form of the $H$-theorem
has been suggested as a way to introduce thermodynamic control
to the system \cite{Karlin,Ansumali}.

From this perspective the goal is to find an equilibrium state
equivalent to the Maxwellian in the continuum which will
similarly maximize entropy. Before the equilibrium can be found
an appropriate $H$ function must be known for a given lattice.
These functions have been constructed in a lattice dependent
fashion in \cite{Karlin}, and $H = -S$ with $S$ from
(\ref{eq:ELBMEntropy}) is an example of a $H$ function
constructed in this way.

One way to implement an ELBM is as a variation on the LBGK, known as
the ELBGK \cite{Ansumali}. In this case $\alpha$ is varied to ensure
a constant entropy condition according to the discrete $H$-theorem.
In general the entropy function is based upon the lattice and cannot
always be found explicitly. However for some examples such as the
simple one dimensional lattice with velocities $\bfv = (-c,0,c)$ and
corresponding populations $\bff = (f_-,f_0,f_+)$ an explicit
Boltzmann style entropy function is known \cite{Karlin}:
\begin{equation}
S(\mathbf{f}) = -f_-\log(f_-) - f_0\log(f_0/4) - f_+\log(f_+).
\label{eq:ELBMEntropy}
\end{equation}
With knowledge of such a function $\alpha$ is found as the non-trivial root of the equation
\begin{equation}
S(\mathbf{f}) = S(\mathbf{f} + \alpha(\mathbf{f}^\ast - \mathbf{f})).
\label{eq:ELBMCondition}
\end{equation}
The trivial root $\alpha = 0$ returns the entropy value of the
original populations. ELBGK then finds the non-trivial $\alpha$ such
that (\ref{eq:ELBMCondition}) holds. This version of the BGK collision
one calls entropic BGK (or EBGK) collision. A solution of
(\ref{eq:ELBMCondition}) must be found at every time step and
lattice site. Entropic equilibria (also derived from the
$H$-theorem) are always used for ELBGK.

\subsection{ELBM algorithm and additional dissipation}

The definition of ELBM for a given entropy equation
(\ref{eq:ELBMCondition}) is incomplete. First of all, it is
possible that the non-trivial solution does not exist.
Moreover, for most of the known entropies (like the perfect
entropy \cite{Karlin}) there always exist such $f$ that the
equation (\ref{eq:ELBMCondition}) for the ELBM collision has no
non-trivial solutions. These $f$ should be sufficiently far
from equilibrium. For completeness, every user of ELBM should
define collisions when the non-trivial root of
(\ref{eq:ELBMCondition}) does not exist. We know and tried two
rules for this situation:
\begin{enumerate}
\item{The most radical approach gives the the Ehrenfest
    rule \cite{GKTO,Brownlee2}: ''if the solution does not
    exist then go to equilibrium", i.e. if the solution
    does not exists then take $\alpha=1$.}
\item{The most gentle solution gives the ''positivity rule"
    \cite{Brownlee2,Li,Tosi,Servan}: to take the maximal
    value of $\alpha$ that guarantees  $f_i +
    \alpha(f_i^\ast - f_i) \geq 0$ for all $i$.}
\end{enumerate}

In general, the Ehrenfest rule prescribes to send the most
non-equilibrium sites to equilibrium and the positivity rule is
applied for any LBM as a recommendation to substitute the
non-positive vectors $\mathbf{f}$ by the closest non-negative
on the interval of the straight line
$[\mathbf{f},\mathbf{f}^\ast]$ that connects $\mathbf{f}$ to
equilibrium. These rules give the examples of the pointwise LBM
limiter and we discuss them separately.

By its nature, the ELBM adds more dissipation than the
positivity rule when the non-trivial root of
(\ref{eq:ELBMCondition}) does not exist. It does not always
keep the entropy balance but increases dissipation for highly
nonequilibrium sites.

\subsection{Numerical method for solution of the ELBM equation}

Another source for additional dissipation in the ELBM may be the
numerical method used for the solution of (\ref{eq:ELBMCondition}). For
the full description of ELBM we have to select a numerical
method for this equation. This method has to have an uniform
accuracy in the wide range of parameters, for all possible
deviation from equilibrium (distribution of these deviations
has ``heavy tails" \cite{BrownleeLimiters2008}).

In order to investigate the stabilization properties of ELBGK
it is necessary to craft a numerical method capable of finding
the non-trivial root in (\ref{eq:ELBMCondition}). In this
section we fix the population vectors $\mathbf{f}$ and
$\mathbf{f}^*$, and are concerned only with this root finding
algorithm. We recast (\ref{eq:ELBMCondition}) as a function of
$\alpha$ only:
\begin{equation}
S_f(\alpha) = S(\mathbf{f} + \alpha(\bffeq - \mathbf{f})) - S(\mathbf{f}).
\label{eq:NewCondition}
\end{equation}
In this setting we attempt to find the non-trivial root $r$ of
(\ref{eq:NewCondition}) such that $S_f(r) = 0$. It should be noted
that as we search for $r$ numerically we should always take care
that the approximation we use is less than $r$ itself. An upper
approximation could result in negative entropy production. A simple
algorithm for finding the roots of a concave function, based on local
quadratic approximations to the target function, has cubic
convergence order. Assume that we are operating in a neighbourhood
$r \in N$, in which $S_f'$ is negative (as well of course $S_f''$ is
negative). At each iteration the new estimate for $r$  is the
greater root of the parabola $P$, the  second order Taylor
polynomial at the current estimate. Analogously to the case for
Newton iteration, the constant in the estimate is the ratio of third
and first derivatives in the interval of iteration:
\begin{eqnarray*}
& |(r-\alpha_{n+1})| \, \le \, C |\alpha_n - r|^3, \\ & \mbox{where}
\;\; C = \frac{1}{6}\sup_{a \in N}|S_f'''(a)| \left/ \inf_{b \in
N}|S_f'(b)| \right. ,
\end{eqnarray*}
where $\alpha_n$ is the evaluation of $r$ on the $n$th iteration.

We use a Newton step to estimate the accuracy of the method at each
iteration: because of the concavity of $S$
\begin{equation}
|\alpha_n - r| \lesssim \left| S_f(\alpha_n)/S_f'(\alpha_n) \right|.
\label{eq:Stopping}
\end{equation}
In fact we use a convergence criteria based not solely on $\alpha$
but on $\alpha||\mathbf{f}^* - \mathbf{f}||$, this has the intuitive
appeal that in the case where the populations are close to the local
equilibrium $\Delta S = S(\mathbf{f}^*) - S(\mathbf{f})$ will be
small and a very precise estimate of $\alpha$ is unnecessary. We
have some freedom in the choice of the norm used and we select
between the standard $L_1$ norm and the entropic norm. The entropic
norm is defined as  $$||\bffeq - \mathbf{f}||_{\bffeq} = -
((\bffeq - \mathbf{f}), \left.D^2
S\right|_{\bffeq}(\bffeq - \mathbf{f})),$$ where $\left.D^2
S\right|_{\bffeq}$ is the second differential of entropy at point
$\bffeq$, and $(x,y)$ is the standard scalar product.

The final root finding algorithm then is beginning with the LBGK
estimate $x_0 = 2$ to iterate using the roots of successive
parabolas. We stop the method at the point,
\begin{equation}
|\alpha_{n} - r| \cdot ||\mathbf{f}^* - \mathbf{f}|| < \epsilon .
\label{eq:ConvergenceTwo}
\end{equation}
To ensure that we use an estimate that is less than the root, at the
point where the method has converged we check the sign of
$S_f(\alpha_n)$. If $S_f(\alpha_n) > 0 $ then we have achieved a lower
estimate, if $S_f(\alpha_n) < 0$ we correct the estimate to the other
side of the root with a double length Newton step,
\begin{equation}
\alpha_n = \alpha_n - 2\frac{S_f(\alpha_n)}{S_f'(\alpha_n)} .
\end{equation}

At each time step before we begin root finding we eliminate all
sites with $\Delta S  < 10^{-15}$. For these sites we make a simple
LBGK step. At such sites we find that round off error in the
calculation of $S_f$ by solution of equation (\ref{eq:ELBMCondition})  can
result in the root of the parabola becoming imaginary. In such cases
a mirror image given by LBGK is effectively indistinct from the
exact ELBGK collision. In the numerical examples given in this work the case where the non-trivial root of the entropy parabola does not exist  was not encountered.

\rubric{4. ENTROPIC FILTERING} \setcounter{section}{4}
\setcounter{subsection}{0} \setcounter{subsubsection}{1}
\label{sec:filtering}

All the specific LBM limiters \cite{BrownleeLimiters2008,Ricot}
are based on a representation of distributions $f$ in the
form:
\begin{equation}\label{separation}
\bff=\bffeq+\|\bff-\bffeq\| \frac{\bff-\bffeq}{\|\bff-\bffeq\|},
\end{equation}
where $\bffeq$ is the corresponding quasiequilibrium
(conditional equilibrium) for given moments $M$, $\bff-\bffeq$
is the nonequilibrium ``part" of the distribution, which is
represented in the form ``norm$\times$direction'' and
$\|f-f^*\|$ is the norm of that nonequilibrium component
(usually this is the entropic norm).

All limiters we use change the norm of the nonequilibrium
component $\bff-\bffeq$, but do not touch its direction or the
quasiequilibrium. In particular, limiters do not change the
macroscopic variables, because moments for $\bff$ and $\bffeq$
coincide. These limiters are transformations of the form
\begin{equation}\label{LimGenF}
\bff \mapsto \bffeq+ \phi \times(\bff-\bffeq)
\end{equation}
with $\phi >0$. If $\bff-\bffeq$ is too big, then the limiter
should decrease its norm.

\begin{figure}
\begin{centering}
\includegraphics[width=0.4\textwidth]{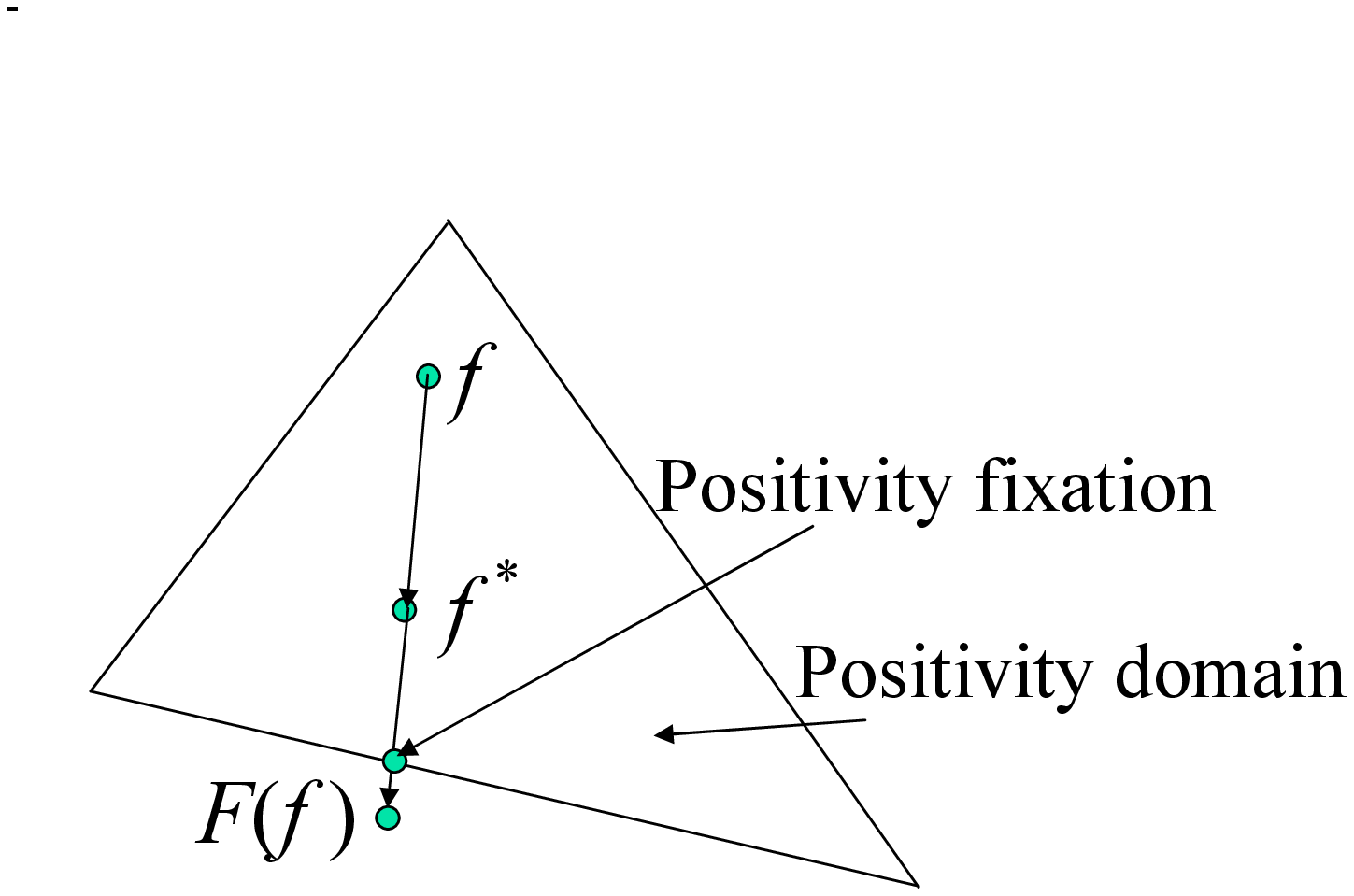}
\caption{\label{fig:PosRule}Positivity rule in action. The motions
stops at the positivity boundary.}
\end{centering}
\end{figure}

For the first example of the realization of this {\em pointwise
filtering} we use the kinetic idea of the {\it positivity
rule}, the prescription is
simple~\cite{Brownlee2,Li,Tosi,Servan}: to substitute
nonpositive $F(\bff)$ by the closest nonnegative state that
belongs to the straight line
\begin{equation}
\label{StrLine}
  \Bigr\{\lambda \bff + (1-\lambda) \bffeq |\: \lambda \in
\mathbb{R}\Bigl\}
\end{equation}
defined by the two points, $\bff$ and the corresponding
quasiequilibrium \ref{fig:PosRule}. This operation is to be applied pointwise, at
points of the lattice where positivity is violated. This technique preserves the
positivity of populations, but can affect the accuracy
of the approximation. This rule is necessary for ELBM
when the positive ``mirror state'' with the same entropy
as $\bff$ does not exists on the straight line~\eqref{StrLine}.

The positivity rule measures the deviation $f-f^*$ by a binary
measure: if all components of this vector $f-f^*$ are
non-negative then it is not too large. If some of them are
negative that this deviation is too large and needs
corrections.

To construct pointwise flux limiters for LBM, based on
dissipation, the entropic approach remains very convenient. The
local nonequilibrium entropy for each site is defined
\begin{equation}\label{locaNEQentropy}
 \Delta S(\bff):= S(\bffeq) - S(\bff).
\end{equation}
The positivity limiter was targeted wherever population densities
became negative. Entropic limiters are targeted wherever
non-equilibrium entropy becomes large.

The first limiter is \emph{Ehrenfests' regularisation}~\cite{Brownlee,Brownlee2}, it
provides ``entropy trimming'': we monitor local deviation of $\bff$
from the corresponding quasiequilibrium, and when $\Delta S(\bff)(\bfx)$
exceeds a pre-specified threshold value $\delta$, perform local
Ehrenfests' steps to the corresponding equilibrium: $\bff \mapsto \bffeq$
at those points.

Not all lattice Boltzmann models are entropic, and an important
question arises: ``how can we create nonequilibrium entropy limiters for
LBM with non-entropic (quasi)equilibria?''. We propose a solution of
this problem based on the discrete Kullback entropy~\cite{Kull}:
\begin{equation}\label{Kul}
S_{\rm K}(\bff)= - \sum_i f_i \ln\biggl(\frac{f_i}{f_i^{\rm eq}}\biggr).
\end{equation}
For entropic quasiequilibria with perfect entropy the discrete
Kullback entropy gives the same $\Delta S$: $-S_{\rm K}(f)= \Delta
S(f)$. Let the discrete entropy have the standard form for an ideal
(perfect) mixture~\cite{Karlin}:
\begin{equation*}
S(\bff)=-\sum_i f_i \ln\biggl(\frac{f_i}{W_i}\biggr).
\end{equation*}
In quadratic approximation,
\begin{equation}\label{KulSq}
-S_{\rm K}(\bff)=\sum_i f_i \ln\biggl(\frac{f_i}{f^{\rm eq}_i}\biggr)
\approx \sum_i\frac{(f_i-f^{\rm eq}_i)^2}{f^{\rm eq}_i}.
\end{equation}
If we define $\bff$ as the conditional entropy maximum for given
$M_j=\sum_k m_{jk} f_k$, then
\begin{equation*}
\ln f^{\rm eq}_k=\sum_j \mu_j m_{jk},
\end{equation*}
where $\mu_j(M)$ are the Lagrange multipliers (or ``potentials'').
For this entropy and conditional equilibrium we find
\begin{equation}\label{DelS}
\Delta S= S(\bffeq)-S(\bff)=\sum_i f_i
\ln\biggl(\frac{f_i}{f^{\rm eq}_i}\biggr)=-S_{\rm K}(\bff),
\end{equation}
if $\bff$ and $\bffeq$ have the same moments,
$m(\bff)=m(\bffeq)$.

In what follows, $\Delta S$ is the Kullback distance $-S_{\rm
K}(\bffeq)$~\eqref{DelS} for general (positive) quasiequilibria
$\bffeq$, or simply $S(\bffeq)-S(\bff)$ for entropic
quasiequilibria (or second approximations for these
quantities~\eqref{KulSq}).

So that the Ehrenfests' steps are not allowed to degrade the
accuracy of LBGK it is pertinent to select the $k$ sites with
highest $\Delta S>\delta$. An a posteriori estimate of added
dissipation could easily be performed by the analysis of entropy
production in the Ehrenfests' steps. Numerical experiments show (see,
e.g.,~\cite{Brownlee,Brownlee2}) that even a small number of such
steps drastically improves stability.

The positivity rule and Ehrenfests' regularisation provide
rare, intense and localised corrections. Of course, it is easy
and also computationally cheap to organise more gentle
transformations with a smooth shift of highly nonequilibrium
states to quasiequilibrium. The following regularisation
transformation with a smooths function $\phi$ distributes its
action smoothly:
\begin{equation}\label{smoothReg}
\bff \mapsto \bffeq + \phi (\Delta S(\bff))(\bff-\bffeq).
\end{equation}
The choice of function $\phi$ is highly ambiguous, for example,
$\phi=1/(1+ \alpha \Delta S^k)$ for some $\alpha>0$ and $k>0$.
There are two significantly different choices: (i)
ensemble-independent $\phi$ (i.e., the value of $\phi$ depends
on local value of $\Delta S$ only) and (ii) ensemble-dependent
$\phi$, for example
\begin{equation}\label{phiExamp}
\phi(\Delta S)=\frac{1+(\Delta S/(\alpha {\rm E} (\Delta
S)))^{k-1/2}} {1+(\Delta S/(\alpha {\rm E} (\Delta S)))^{k}},
\end{equation}
where ${\rm E}( \Delta S)$ is the average value of $\Delta S$
in the computational area, $k\geq 1$, and $\alpha \gtrsim 1$.
For small $\Delta S$, $\phi(\Delta S) \approx 1$ and for
$\Delta S \gg \alpha {\rm E} (\Delta S)$ it tends to
$\sqrt{\alpha {\rm E} (\Delta S) / \Delta S}$. It is easy to
select an ensemble-dependent $\phi$ with control of total
additional dissipation.

\subsection{Monotonic and double monotonic limiters}

Two monotonicity properties are important in the theory of
nonequilibrium entropy limiters:
\begin{enumerate}
\item A limiter should move the distribution to
    equilibrium: in all cases of~\eqref{LimGenF} $0 \leq
    \phi \leq 1$. This is the \emph{ dissipativity}
    condition which means that limiters never produce
    negative entropy.
\item A limiter should not change the order of states on
    the line: if for two distributions with the same
    moments, $\bff$ and $\bff '$,
    $\bff'-\bffeq=x(\bff-\bffeq)$ and $\Delta S (\bff) >
    \Delta S (\bff')$ before the limiter transformation,
    then the same inequality should hold after the limiter
    transformation too. For example, for the
    limiter~\eqref{smoothReg} it means that $\Delta
    S(\bffeq +x \phi (\Delta
    S(\bffeq+x(\bff-\bffeq))(\bff-\bffeq))$ is a
    monotonically increasing function of $x>0$.
\end{enumerate}
In quadratic approximation,
\begin{align*}
\Delta S(\bffeq +x(\bff-\bffeq)) &= x^2 \Delta S(\bff),\\
\Delta S(\bffeq +x \phi (\Delta S(\bffeq +x(\bff-\bffeq))(\bff-\bffeq)) &= x^2
\phi^2(x^2 \Delta S(\bff)),
\end{align*}
and the second monotonicity condition transforms into the
following requirement: $y \phi (y^2 s)$ is a monotonically
increasing (not decreasing) function of $y>0$ for any $s>0$.

If a limiter satisfies both monotonicity conditions, we call it
``double monotonic''. For example, Ehrenfests' regularisation
satisfies the first monotonicity condition, but violates the
second one. The limiter~\eqref{phiExamp} violates the first
condition for small $\Delta S$, but is dissipative and
satisfies the second one in quadratic approximation for large
$\Delta S$. The limiter with $\phi=1/(1+ \alpha \Delta S^k)$
always satisfies the first monotonicity condition, violates the
second if $k > 1/2$, and is double monotonic (in quadratic
approximation for the second condition), if $0 < k \leq 1/2$.
The threshold limiter~\eqref{entrHom} is also double monotonic.

For smooth functions, the condition of double monotonicity (in
quadratic approximation) is equivalent to the system of
differential inequalities:
\begin{equation*}
\begin{split}
\phi(x) + 2 x \phi ' (x) & \geq 0;\\ \phi' (x) &\leq 0.
\end{split}
\end{equation*}
The initial condition $\phi(0)=1$ means that in the limit
$\Delta S \to 0$ limiters do not affect the flow. Following
these inequalities we can write: $2x \phi '(x) = - \eta (x)
\phi(x) $, where $0\leq \eta (x) \leq 1$. The solution of these
inequalities with initial condition is
\begin{equation}\label{DoublMonGen}
\phi(x)=\exp\biggl(-\frac{1}{2} \int_0^x \frac{\eta (\chi)}{\chi} \D
\chi \biggr),
\end{equation}
if the integral on the right-hand side exists. This is a
general solution for double monotonic limiters (in the second
approximation for entropy). If $\eta (x)$ is the Heaviside step
function, $\eta (x) = H(x-\Delta S_{\rm t})$ with threshold
value $\Delta S_{\rm t}$, then the general solution
(\ref{DoublMonGen}) gives us the threshold limiter. If, for
example, $\eta (x)= x^k / (\Delta S_{\rm t}^k + x^k)$, then
\begin{equation}\label{DoublMonSpec}
\phi(x)=\left(1 + \frac{x^k}{\Delta S_{\rm
t}^k}\right)^{-\frac{1}{2k}}.
\end{equation}
This special form of limiter function is attractive because for
small $x$ it gives
\begin{equation*}
\phi(x)= 1-\frac{1}{2k}\frac{x^k}{\Delta S_{\rm t}^k} +o(x^k).
\end{equation*}
Thus, the limiter does not affect the motion up to the
$(k+1)$st order, and the macroscopic equations coincide with
the macroscopic equations for LBM without limiters up to the
$(k+1)$st order in powers of deviation from quasiequilibrium.
Furthermore, for large $x$ we get the $k$th order approximation
to the threshold limiter \eqref{entrHom}:
\begin{equation*}
\phi(x)=\sqrt{\frac{\Delta S_{\rm t}}{x}} + o(x^{-k}).
\end{equation*}

Of course, it is not forbidden to use any type of limiters
under the local and global control of dissipation, but double
monotonic limiters provide some natural properties
automatically, without additional care.

\subsection{Monitoring total dissipation \label{Monitor}}
For given $\beta$, the entropy production in one LBGK step in
quadratic approximation for $\Delta S$ is:
\begin{equation*}
\delta_{\rm LBGK} S \approx [1-(2\beta-1)^2] \sum_{x} \Delta S
(\bfx),
\end{equation*}
where $\bfx$ is the grid point, $\Delta S (\bfx)$ is
nonequilibrium entropy~\eqref{locaNEQentropy} at point $\bfx$,
$\delta_{\rm LBGK} S$ is the total entropy production in a
single LBGK step. It would be desirable if the total entropy
production for the limiter $\delta_{\rm lim} S$ was small
relative to $\delta_{\rm LBGK} S$:
\begin{equation}\label{LimEntLim}
\delta_{\rm lim} S  < \delta_0 \delta_{\rm LBGK} S.
\end{equation}

A simple ensemble-dependent limiter (perhaps, the simplest one)
for a given $\delta_0$ operates as follows. Let us collect the
histogram of the $\Delta S(\bfx)$ distribution, and estimate
the distribution density, $p(\Delta S)$. We have to estimate a
value $\Delta S_0$ that satisfies the following equation:
\begin{equation}\label{EntThresh}
\int_{\Delta S_0}^{\infty} p(\Delta S) (\Delta S - \Delta S_0) \,
\D \Delta S = \delta_0 [1-(2\beta-1)^2] \int_0^{\infty} p(\Delta
S) \Delta S \, \D \Delta S.
\end{equation}
In order not to affect distributions with a small expectation of
$\Delta S$, we choose a threshold $\Delta S_{\rm t}= \max \{ \Delta
S_0, \delta\}$, where $\delta$ is some predefined value (as in the
Ehrenfests' regularization). For states at sites with $\Delta S \geq
\Delta S_{\rm t} $ we provide homothety with equilibrium center
 $\bffeq$ and coefficient $\sqrt{\Delta S_{\rm t} /
\Delta S}$ (in quadratic approximation  for nonequilibrium entropy):
\begin{equation}\label{entrHom}
\bff(\bfx) \mapsto \bffeq(\bfx) + \sqrt{\frac{\Delta S_{\rm t}}{\Delta S}}
(\bff(\bfx)- \bffeq(\bfx)) .
\end{equation}

To avoid the change of accuracy order ``on average'', the number of
sites with this step should be $\leq \mathcal{O}(Nh/L)$ where $N$ is
the total number of sites, $h$ is the step of the space
discretization and $L$ is the macroscopic characteristic length. But
this rough estimate of accuracy across the system might be destroyed by
a concentration of Ehrenfests' steps in the most nonequilibrium areas,
for example, in boundary layers. In that case, instead of the total
number of sites $N$ in $\mathcal{O}(Nh/L)$ we should take the number
of sites in a specific region. The effects of such concentration could be
analysed a posteriori.

\subsection{Median entropy filter}
The Ehrenfest step described above provides pointwise correction of
nonequilibrium entropy at the ``most nonequilibrium'' points. Due to
the pointwise nature, the technique does not introduce any
nonisotropic effects, and provides some other benefits. But if we
involve local structure, we can correct local non-monotone
irregularities without touching regular fragments. For example, we
can discuss monotone increase or decrease of nonequilibrium entropy
as regular fragments and concentrate our efforts on reduction of
``speckle noise'' or ``salt and pepper noise''. This approach allows
us to use the accessible resource of entropy
change~\eqref{LimEntLim} more thriftily. Salt and pepper noise is a
form of noise typically observed in images. It represents itself as
randomly occurring white (maximal brightness) and black pixels. For
this kind of noise, conventional low-pass filtering, e.g., mean
filtering or Gaussian smoothing is unsuccessful because the
perturbed pixel value can vary significantly both from the original
and mean value. For this type of noise, \emph{median filtering} is a
common and effective noise reduction method. Median filtering is a
common step in image processing~\cite{Pratt} for the smoothing of
signals and the suppression of impulse noise with preservation of
edges.

The median is a more robust average than the mean (or the weighted
mean) and so a single very unrepresentative value in a
neighbourhood will not affect the median value significantly.
Hence, we suppose that the median entropy filter will work better
than entropy convolution filters.

For the nonequilibrium entropy field, the median filter considers
each site in turn and looks at its nearby neighbours. It replaces
the nonequilibrium entropy value $\Delta S$ at the point with the
median of those values $\Delta S_{\rm med}$, then updates $f$ by
the transformation~\eqref{entrHom} with the homothety coefficient
$\sqrt{\Delta S_{\rm med}/{\Delta S}}$. The median, $\Delta S_{\rm
med}$, is calculated by first sorting all the values from the
surrounding neighbourhood into numerical order and then replacing
that being considered with the middle value. For example, if a
point has 3 nearest neighbours including itself, then after sorting
we have 3 values $\Delta S$: $\Delta S_1 \leq \Delta S_2 \leq
\Delta S_3$. The median value is $\Delta S_{\rm med}=\Delta S_2$.
For 9 nearest neighbours (including itself) we have after sorting
$\Delta S_{\rm med}=\Delta S_5$. For 27 nearest neighbours  $\Delta
S_{\rm med}=\Delta S_{14}$.

We accept only dissipative corrections (those resulting in a
decrease of $\Delta S$, $\Delta S_{\rm med}< \Delta S$) because of
the second law of thermodynamics. The analogue of~\eqref{EntThresh}
is also useful for the acceptance of the most significant corrections.
In ``salt and pepper" terms, we correct the salt (where $\Delta S$
exceeds the median value) and do not touch the pepper.

\subsection{General nonlocal filters}

The separation of $\bff$ in equilibrium and nonequilibrium
parts (\ref{separation}) allows one to use any nonlocal
filtering procedure. Let $\bff^{\rm neq}=\bff-\bffeq$. The
values of moments for $\bff$ and $\bffeq$ coincide, hence we
can apply any transformation of the form $$\bff^{\rm neq}(\bfx)
\mapsto \sum_{y} d(\bfy)\bff^{\rm neq}(\bfx+\bfy)$$ for any
family of vectors $\bfy$ that shift the grid into itself and
any coefficients $d(\bfy)$. If we apply this transformation,
the macroscopic variables do not change but their time
derivatives may change. We can control the values of some
higher moments in order not to perturb significantly some
macroscopic parameters, the shear viscosity, for example
\cite{Dellar2001}. Several local (but not pointwise) filters of
this type have been proposed and tested recently \cite{Ricot}.

\rubric{5. MULTIPLE RELAXATION TIMES} \setcounter{section}{5}
\setcounter{subsection}{0} \setcounter{subsubsection}{1} The
MRT lattice Boltzmann system~\cite{Lallemand,Luo,Dellar}
generalizes the BGK collision into a more general linear
transformation of the population functions,
\begin{equation}
F(\bff) = \bff + A(\bffeq - \bff),
\end{equation}
where $A$ is a square matrix of size $m$. The use of this more
general operator allows more different parameters to be used within
the collision, to manipulate different physical properties, or for
stability purposes.

To facilitate this a change of basis matrix can be used to switch
the space of the collision to the moment space. Since the moment
space of the system may be several dimensions smaller than the
population space, to complete the basis linear combinations of
higher order polynomials of the discrete velocity vectors may be
used. For our later experiments we will use the D2Q9 system, we
should select a particular enumeration of the discrete velocity
vectors for the system, the zero velocity is numbered one and the
positive x velocity is numbered 2, the remainder are numbered
clockwise from this system,
\begin{equation}
\begin{array}{ccc}
7 & 8 & 9 \\
6 & 1 & 2 \\
5 & 4 & 3
\end{array}.
\end{equation}
The change of basis matrix given in some of the literature
\cite{Lallemand} is chosen to represent specific macroscopic
quantities and in our velocity system is as follows,
\begin{equation}
M_1 = \left( \begin{array}{c c c c c c c c c}
1 & 1 & 1 & 1 & 1 & 1 & 1 & 1 & 1\\
-4 & -1 & 2 & -1 & 2 & -1 & 2 & -1 & 2\\
4 & -2 & 1 & -2 & 1 & -2 & 1 & -2 & 1\\
0 & 1 & 1 & 0 & -1 & -1 & -1 & 0 & 1\\
0 & -2 & 1 & 0  & -1 & 2 & -1 & 0 & 1\\
0 & 0 & -1 & -1 & -1 & 0 & 1 & 1 & 1\\
0 & 0 & -1 & 2 & -1 & 0 & 1 & -2 & 1 \\
0 & 1 & 0 & -1 & 0 & 1 & 0 & -1 & 0\\
0 & 0 & -1 & 0 & 1 & 0 & -1 & 0 & 1
\end{array}\right)
\end{equation}
As an alternative we could complete the basis simply using higher powers of the velocity vectors,
the basis would be
$1,v_x,v_y,v_x^2+v_y^2,v_x^2-v_y^2,v_xv_y,v_x^2v_y,v_xv_y^2,v_x^2v_y^2$, in our
velocity system then this change of basis matrix is as follows,
\begin{equation}
M_2 = \left( \begin{array}{c c c c c c c c c}
1 & 1 & 1 & 1 & 1 & 1 & 1 & 1 & 1\\
0 & 1 & 0 & -1 & 0 & 1 & -1 & -1 & 1\\
0 & 0 & 1 & 0 & -1 & 1 & 1 & -1 & -1\\
0 & 1 & 1 & 1  & 1 & 2 & 2 & 2 & 2\\
0 & 1 & -1 & 1 & -1 & 0 & 0 & 0 & 0\\
0 & 0 & 0 & 0 & 0 & 1 & -1 & 1 & -1 \\
0 & 0 & 0 & 0 & 0 & 1 & -1 & -1 & 1\\
0 & 0 & 0 & 0 & 0 & 1 & 1 & -1 & -1\\
0 & 0 & 0 & 0 & 0 & 1 & 1 & 1 & 1
\end{array}\right)
\end{equation}
When utilized properly any basis should be equivalent (although with
different rates). In any case in this system there are 3 conserved
moments and 9 population functions. Altogether then there are 6
degrees of freedom in relaxation in this system. We need some of
these degrees of freedom to implement hydrodynamic rates such as
shear viscosity or to force isotropy and some are `spare' and can be
manipulated to improve accuracy or stability. Typically these spare
relaxation modes are sent closer to equilibrium than the standard
BGK relaxation rate. This is in effect an additional contraction in
the finite dimensional non-equilibrium population function space,
corresponding to an increase in dissipation.

Once in moment space we can apply a diagonal relaxation matrix $C_1$
to the populations and then the inverse moment transformation matrix
$M_1^{-1}$ to switch back into population space, altogether $A_1  =
M_1^{-1}C_1M_1$. If we use the standard athermal polynomial
equilibria then three entries on the diagonal of $C_1$ are actually
not important as the moments will be automatically conserved since
$m(\bff) = m(\bffeq)$, for simplicity we set them equal to 0 or 1 to
reduce the complexity of the terms in the collision matrix. There
are 6 more parameters on the diagonal matrix $C$ which we can set.
Three of these correspond to second order moments, one each is
required for shear and bulk viscosity which are called $s_e$ and
$s_\nu$ respectively and one for isotropy. Two correspond to third
order moments, one gives a relaxation rate $s_q$ and again one is
needed for isotropy. Finally one is used to give a relaxation rate
$s_\epsilon$ for the single fourth order moment. We have then in
total four relaxation parameters which appear on the diagonal matrix
in the following form:
\begin{equation}
C_1 = \left( \begin{array}{ c c c c c c c c c}
1 & 0 & 0 & 0 & 0 & 0 & 0 & 0 & 0\\
0 & s_e & 0 & 0 & 0 & 0 & 0 & 0 & 0\\
0 & 0 & s_\epsilon & 0 & 0 & 0 & 0 & 0 & 0\\
0 & 0 & 0 & 1 & 0 & 0 & 0 & 0 & 0\\
0 & 0 & 0 & 0 & s_q & 0 & 0 & 0 & 0\\
0 & 0 & 0 & 0 & 0 & 1 & 0 & 0 & 0\\
0 & 0 & 0 & 0 & 0 & 0 & s_q & 0 & 0\\
0 & 0 & 0 & 0 & 0 & 0 & 0 & s_\nu & 0\\
0 & 0 & 0 & 0 & 0 & 0 & 0 & 0 & s_\nu
\end{array}\right).
\end{equation}
Apart from the parameter $s_\nu$ which is used to control shear
viscosity, in an incompressible system the other properties can be
varied to improve accuracy and stability. In particular, there
exists a variant of MRT known as TRT (two relaxation time)
\cite{Ginzburg} where the relaxation rates $s_e,s_\epsilon$ are made
equal to $s_\nu$. In a system with boundaries the final rate is
calculated $s_q = 8(2=s_\nu)/(8-s_\nu)$, this is done, in
particular, to combat numerical slip on the boundaries of the
system.

We should say that in some of the literature regarding MRT the
equilibrium is actually built in moment space, that is the collision
operation would be written,
\begin{equation}
F(\bff) = \bff + M_1^{-1}C_1(\mathbf{m}^{\mathrm{eq}} - M_1 \bff).
\end{equation}
This could be done to increase efficiency, depending on the
implementation, however each moment equilibrium
$\mathbf{m}^{\mathrm{eq}}$ has an equivalent population space
equilibrium $\bffeq = M_1^{-1} \mathbf{m}^{\mathrm{eq}} $, so the
results of implementing either system should be the same up to
rounding error.

We can also conceive of using an MRT type collision as a limiter,
that is to apply it only on a small number of points on the lattice
where non equilibrium entropy passes a certain threshold. This
answers a criticism of the single relaxation time limiters that they
fail to preserve dissipation on physical modes. As well as using the
standard MRT form given above we can build an MRT limiter using the
alternative change of basis matrix $M_2$.

The limiter in this case is based on the idea of sending every mode
except shear viscosity directly to equilibrium again the complete
relaxation matrix is given by $A_2 = M_2^{-1}C_2M_2$ where,
\begin{equation}
C_2 = \left( \begin{array}{ c c c c c c c c c}
1 & 0 & 0 & 0 & 0 & 0 & 0 & 0 & 0\\
0 & 1 & 0 & 0 & 0 & 0 & 0 & 0 & 0\\
0 & 0 & 1 & 0 & 0 & 0 & 0 & 0 & 0\\
0 & 0 & 0 & 1 & 0 & 0 & 0 & 0 & 0\\
0 & 0 & 0 & 0 & s_\nu & 0 & 0 & 0 & 0\\
0 & 0 & 0 & 0 & 0 & s_\nu & 0 & 0 & 0\\
0 & 0 & 0 & 0 & 0 & 0 & 1 & 0 & 0\\
0 & 0 & 0 & 0 & 0 & 0 & 0 & 1 & 0\\
0 & 0 & 0 & 0 & 0 & 0 & 0 & 0 & 1
\end{array}\right).
\end{equation}
This could be considered a very {\it aggressive} form of the MRT
which maximizes regularization on every mode except shear viscosity
and would not be appropriate for general use in a system, especially
as most systems violate the incompressibility assumption and hence
bulk viscosity is not small. The advantage of using the different
change of basis matrix is that the complete collision matrix $A_2$
is relatively sparse with just twelve off diagonal elements and
hence is easy to implement and not too expensive to compute with.

\rubric{6. 1D SHOCK TUBE}
\setcounter{section}{1}

A standard experiment for the testing of LBMs is the one-dimensional
shock tube problem. The lattice velocities used are $\mathbf{v} =
(-1,0,1)$, so that space shifts of the velocities give lattice sites
separated by the unit distance. 800 lattice sites are used and are
initialized with the density distribution
\begin{displaymath}
 \rho(x) =  \left\{
 \begin{array}{ll}
 1, & \;\;\;\;1\leq x \leq 400, \\
 0.5, &\;\;\;\;401 \leq x \leq 800.
 \end{array}
 \right.
\end{displaymath}
Initially all velocities are set to zero. We compare the ELBGK
equipped with the parabola based root finding algorithm using the
entropic norm with the standard LBGK method using both standard
polynomial and entropic equilibria. The polynomial equilibria are
given in \cite{Benzi,Succi}:
\begin{displaymath}
\begin{split}
&f_-^* = \frac{\rho}{6}\left(1 - 3u + 3u^2\right), \; \; f_0^* =
\frac{2\rho}{3}\left(1 - \frac{3u^2}{2}\right), \\ & f_+^* =
\frac{\rho}{6}\left(1 + 3u + 3u^2\right).
\end{split}
\label{eq:1Dpoly}
\end{displaymath}
The entropic equilibria also used by the ELBGK are available
explicitly as the maximum of the entropy function
(\ref{eq:ELBMEntropy}),
\begin{displaymath}
\begin{split}
&f_-^* = \frac{\rho}{6}(-3u - 1 + 2\sqrt{1\!+\!3u^2}), \; \; f_0^* =
\frac{2\rho}{3}(2 - \sqrt{1\!+\!3u^2}), \\ &  f_+^* =
\frac{\rho}{6}(3u - 1 + 2\sqrt{1\!+\!3u^2}).
\end{split}
\end{displaymath}
Now following the prescription fromm Sec. \ref{sec:bac} the
governing equations for the simulation are
\begin{displaymath}
\begin{split}
 &f_-(x,t\!+\!1)\! =\! f_-(x\!+\!1,t)\! +\! \alpha\beta(f_-^*(x\!+\!1,t)\! - \!f_-(x\!+\!1,t)), \\
 &f_0(x,t\!+\!1)\! =\! f_0(x,t)\! +\! \alpha\beta(f_0^*(x,t) - f_0(x,t)), \\
 &f_+(x,t\!+\!1)\! =\! f_+(x\!-\!1,t)\! +\! \alpha\beta(f_+^*(x\!-\!1,t)\! -\! f_+(x\!-\!1,t)).
\end{split}
\end{displaymath}
\begin{figure}[h!]
\centering
\includegraphics[width=160 mm]{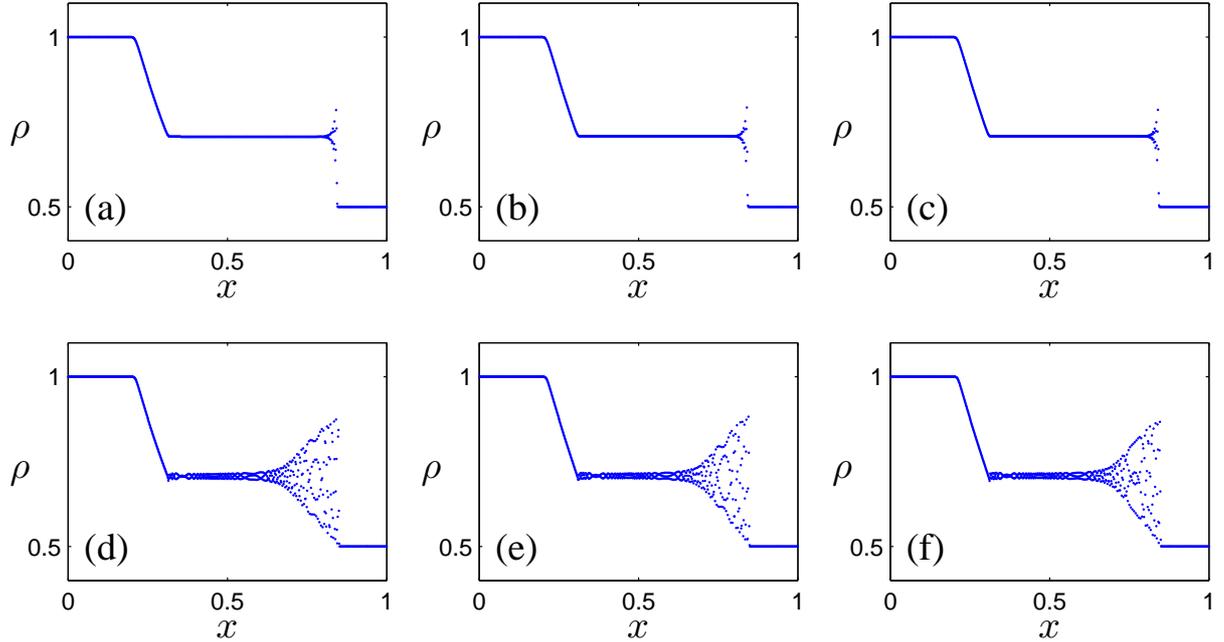}
\caption{Density profile of the simulation of the shock tube problem
following 400 time steps using (\textbf{a}) LBGK with polynomial
equilibria [$\nu = (1/3)\cdot10^{-1}$]; (\textbf{b}) LBGK with
entropic equilibria [$\nu = (1/3)\cdot10^{-1}$]; (\textbf{c}) ELBGK
[$\nu = (1/3)\cdot10^{-1}$]; (\textbf{d}) LBGK with polynomial
equilibria [$\nu = 10^{-9}$]; (\textbf{e}) LBGK with entropic
equilibria [$\nu = 10^{-9}$]; (\textbf{f}) ELBGK [$\nu =
10^{-9}$].\label{fig:ELBMResults}}
\end{figure}
From this experiment we observe no benefit in terms of
regularization in using the ELBGK rather than the standard LBGK
method (Fig. \ref{fig:ELBMResults}). In both the medium and low
viscosity regimes ELBGK does not supress the spurious
oscillations found in the standard LBGK method. The
observation is in full agreement with the Tadmor and Zhong
\cite{Tadmor2006} experiments for schemes with precise entropy
balance.

Entropy balance gives a nice additional possibility to monitor
the accuracy and the basic physics but does not give an
omnipotent tool for regularization.

\rubric{7. 2D SHEAR DECAY} \setcounter{section}{7}
\setcounter{subsection}{0} \setcounter{subsubsection}{1}

For the second test we use a simple test proposed to measure the
observable viscosity of a lattice Boltzmann simulation to validate
the shear viscosity production of the MRT models. We take the 2D
isothermal nine-velocity model with standard polynomial equilibria.
Our computational domain will a square which we discretize with $L+1
\times L+1$ uniformly spaced points and periodic boundary
conditions. The initial condition is $\rho(x,y)=1, u_x(x, y)=0$ and
$u_y(x, y)=u_0 sin(2\pi x/L)$, with $u_0=0.05$. The exact velocity
solution to this problem is an exponential decay of the initial
condition: $u_x(x, y , t)=0, u_y(x, y , t) =u_0 \exp(-\lambda u_0t /
{\rm Re} L)\sin(2 \pi x/L)$, where $\lambda$ is some constant and
$Re = u_0 L/\nu$ is the Reynolds number of the flow. Here, $\nu$ is
the theoretical shear viscosity of the fluid due to the relaxation
parameters of the collision operation.

Now, we simulate the flow over $L/u_0$ time steps and measure the
constant $\lambda$ from the numerical solution. We do this for LBGK,
the `aggressive' MRT with collision matrix $A_2$ and the MRT system
with collision matrix $A_1$ with additional parameters $s_e = 1.64,
s_\epsilon = 1.54, s_q = 3(2-s_\nu)/(3-s_\nu)$ \cite{Lallemand}. The
shear viscosity relaxation parameter $s_\nu$ is varied to give
different viscosities and therefore Reynolds numbers  for $L=50$ and
for $L=100$.

\begin{figure}[h!]
\includegraphics[width=140 mm]{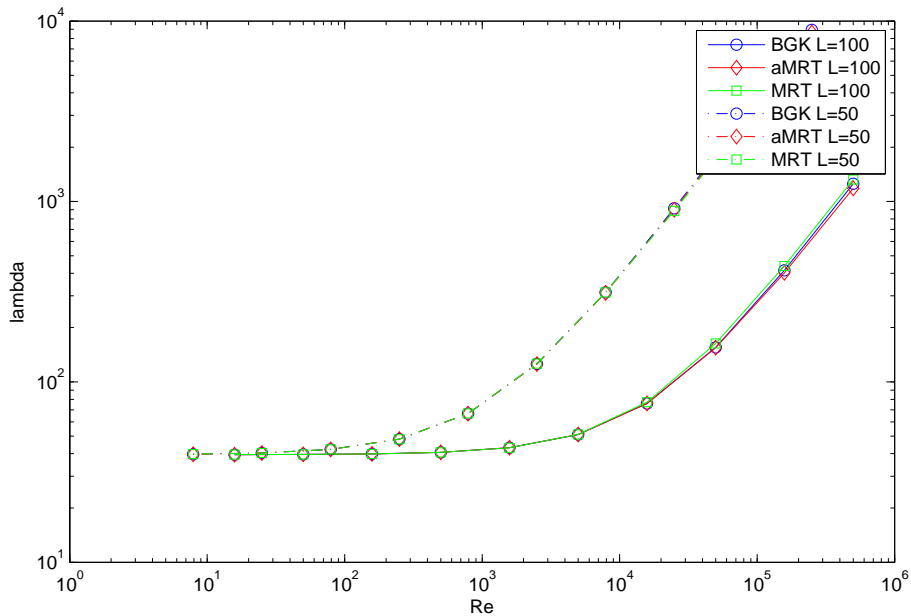}
\caption{Observed $\lambda$ in the shear decay experiment \label{fig:ShearResults}}
\end{figure}

From Figure \ref{fig:ShearResults} it can be seen that in all the
systems, for increasing theoretical Reynolds numbers (decreasing
viscosity coefficient) following a certain point numerical
dissipation due to the lattice begins growing. The most important
observation from this system is that our MRT models do indeed
produce shear viscosity at almost the same rate as the BGK model.

In particular, the bulk viscosity in this system is zero, so all
dissipation is given by shear viscosity or higher order modes. Since
the space derivatives of the velocity modes are well bounded, as is
the magnitude of the velocity itself, the proper asymptotic decay of
higher order modes is observed and the varying higher order
relaxation coefficients have only a very marginal effect.

We should reiterate that we while we use the `aggressive' MRT across
the system, this is only appropriate as bulk viscosity is zero and
in fact the selection of the bulk viscosity coefficient makes no
difference in this example. This example is a special case in this
regard.

\rubric{8. LID DRIVEN CAVITY}
\setcounter{section}{8}
\setcounter{subsection}{0}
\setcounter{subsubsection}{1}

\subsection{Stability}

Our next 2D example is the benchmark 2D lid driven cavity. In this
case this is a square system of side length 129. Bounce back
boundary conditions are used and the top boundary imposes a constant
velocity of $u_{\rm wall} = 0.1$. For a variety of Reynolds numbers
we run experiments for up to 10000000 time steps and check which
methods have remained stable up until that time step.

The methods which we test are the standard BGK system, the BGK
system equipped with Ehrenfest steps, the BGK system equipped with
the MRT limiter, the TRT system, an MRT system with the TRT
relaxation rate for the third order moment and the other rates $s_e
= 1.64, s_\epsilon = 1.54$ and finally an MRT system which we call
MRT1 with the rates $s_q =1.9 s_e = 1.64, s_\epsilon = 1.54$
\cite{Luo}. In each case of the system equipped with limiters the
maximum number of sites where the limiter is used is 9.

All methods are equipped with the standard 2nd order compressible
quasi-equilibrium, which is available as the product of the 1D
equilibria \ref{eq:1Dpoly}.

When calculating the stream functions of the final states of these
simulations we use Simpson integration in first the x and then y
directions.

Additionally we measure Enstrophy $\mathcal{E}$ in each system over
time. Enstrophy is calculated as the sum of vorticity squared across
the system, normalized by the number of lattice sites. This
statistic is useful as vorticity is theoretically only dissipated
due to shear viscosity, at the same time in the lid driven system
vorticity is produced by the moving boundary. For these systems
$\mathcal{E}$ becomes constant as the vorticity field becomes
steady. The value of this constant indicates where the {\it balance}
between dissipation and production of vorticity is found. The lower
the final value of $\mathcal{E}$ the more dissipation produced in
the system.

\begin{figure}[h!]
\centering
\includegraphics[width=\textwidth]{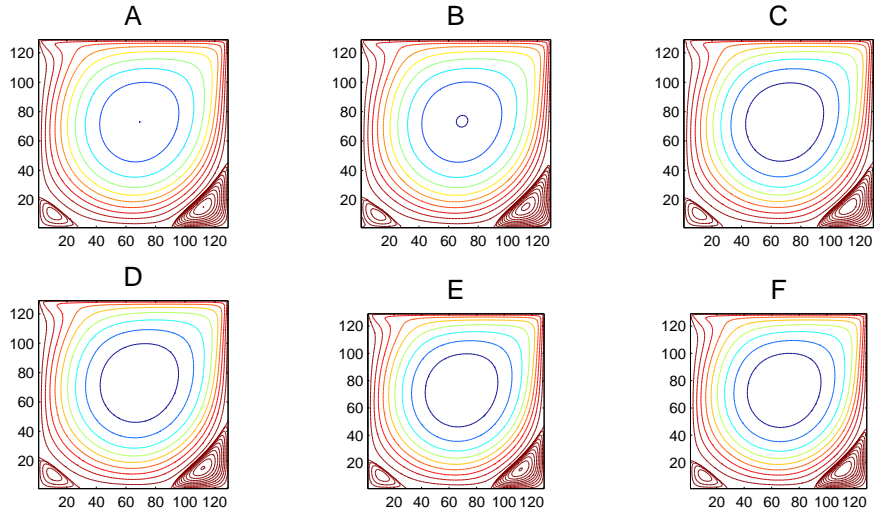}
\caption{Contour plots of stream functions of A: BGK, B: BGK +
Ehrenfest Steps, C: BGK + MRT Limiter, D: TRT, E: MRT, F: MRT1
following 10000000 time steps at Re1000.\label{fig:Re1000conts}}
\end{figure}

\begin{figure}[h!]
\centering
\includegraphics[width=0.5\textwidth]{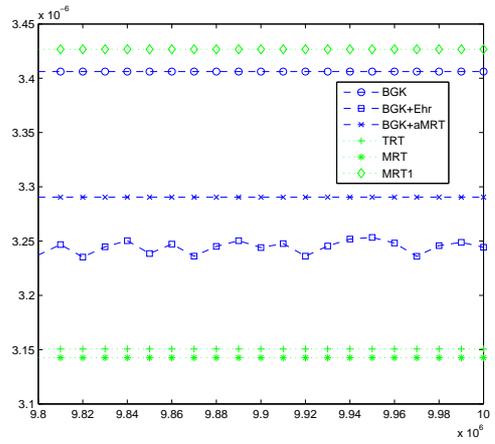}
\caption{Enstrophy in the Re1000 systems during the final $2 \cdot
10^5$ time steps\label{fig:Re1000ens}}
\end{figure}

All of these systems are stable for Re1000, the contour plots of the
final state are given in Figure \ref{fig:Re1000conts} and there
appears only small differences. We calculate the average enstrophy
in each system and plot it as a function of time in Figure
\ref{fig:Re1000ens}. We can see that in the different systems that
enstrophy and hence dissipation varies. Compared with the BGK system
all the other systems except MRT1 exhibit a lower level of enstrophy
indicating a higher rate of dissipation. For MRT1 the fixed
relaxation rate of the third order mode is actually less dissipative
than the BGK relaxation rate for this Reynolds number, hence the
increased enstrophy. An artifact of using the pointwise filtering
techniques is that they introduce small scale local oscillations in
the modes that they regularize, therefore the system seems not to be
asymptotically stable. This might be remedied by increasing the
threshold of $\Delta S$ below which no regularization is performed.
Nevertheless in these experiments after sufficient time the
enstrophy values remain within a small enough boundary for the
results to be useful.

\begin{figure}[h!]
\centering
\includegraphics[width=\textwidth]{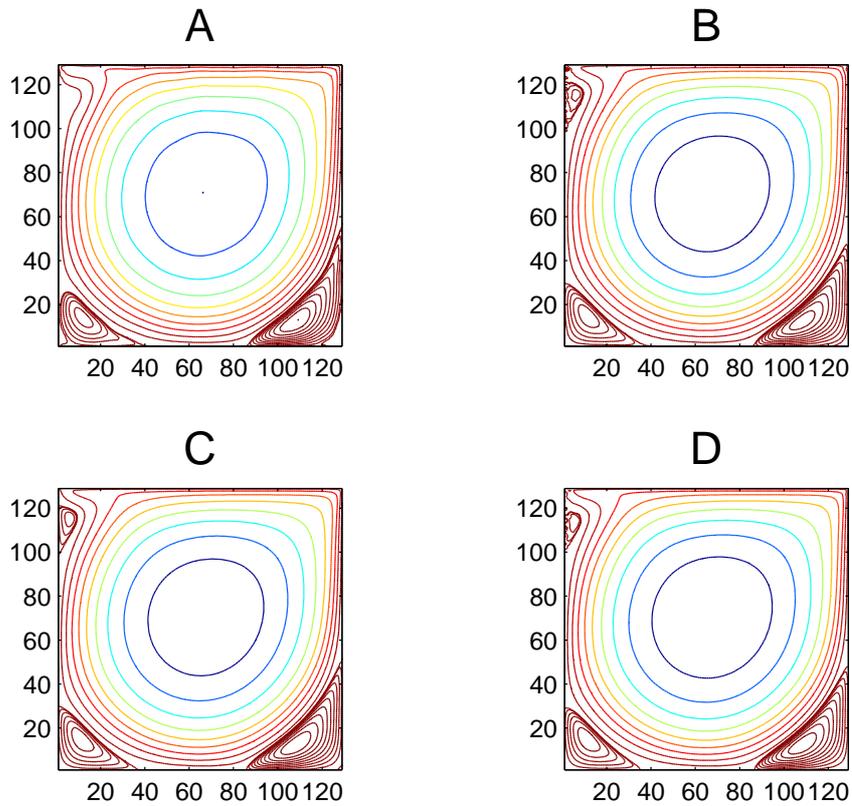}
\caption{Contour plots of stream functions of A: BGK + Ehrenfest
Steps, B: BGK + MRT Limiter, C: TRT, D: MRT1,following 10000000 time
steps at Re2500. \label{fig:Re2500conts}}
\end{figure}

\begin{figure}[h!]
\centering
\includegraphics[width=0.5\textwidth]{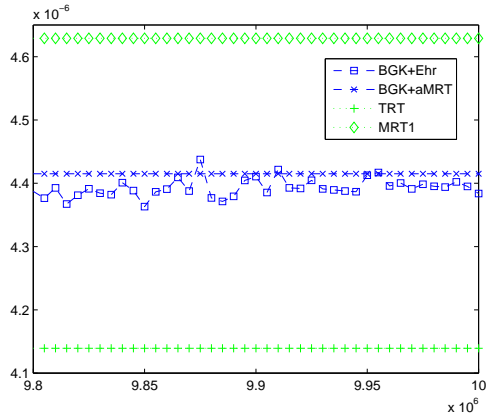}
\caption{Enstrophy in the Re2500 systems during the final $2\cdot
10^5$ time steps\label{fig:Re2500ens}}
\end{figure}

The next Reynolds number we choose is Re2500. Only 4 of the original
6 systems complete the full number of time steps for this Reynolds
number, the contour plots of the final stream functions are given in
Figure \ref{fig:Re2500conts}. Of the systems which did not complete
the simulation it should be said that the MRT system survived a few
10s of thousands of time steps while the BGK system diverged almost
immediately, indicating that it does provide stability benefits
which are not apparent at the coarse granularity of Reynolds numbers
used in this study. One feature to observe in the stream function
plots is the absence of an upper left vortex in the Ehrenfest
limiter. This system selects the "most non-equilibrium" sites to
apply the filter. These typically occur near the corners of the
moving lid. It seems here that the local increase in shear viscosity
is enough to prevent this vortex forming. This problem does not seem
to affect the MRT limiter which preserves the correct production of
shear viscosity.

Again we check the enstrophy of the systems and give the results for
the final timesteps in Figure \ref{fig:Re2500ens}. Due to the
failure of the BGK system to complete this simulation there is no
"standard" result to compare the improved methods with. The
surviving methods maintain their relative positions with respect to
enstrophy production.

\begin{figure}[h!]
\centering
\includegraphics[width=0.8\textwidth]{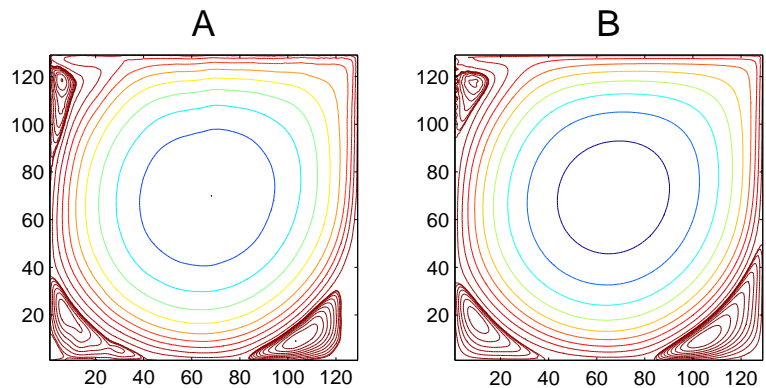}
\caption{Contour plots of stream functions of A: BGK + Ehrenfest
Steps, B: MRT1,following 10000000 time steps at Re5000.
\label{fig:Re5000conts}}
\end{figure}

\begin{figure}[h!]
\centering
\includegraphics[width=0.5\textwidth]{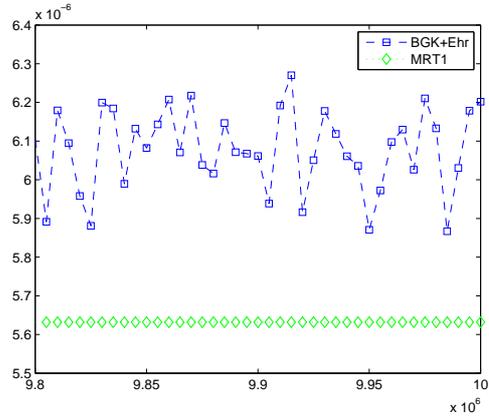}
\caption{Enstrophy in the Re5000 systems during the final $2\cdot
10^5$ time steps\label{fig:Re5000ens}}
\end{figure}

For the theoretical Reynolds number of 5000 only two systems remain,
their streamfunction plots are given in Figure
\ref{fig:Re5000conts}. At this Reynolds number the upper left vortex
has appeared in the Ehrenfest limited simulation, however a new
discrepancy has arisen. The lower right corner exhibits a very low
level of streaming.

In Figure \ref{fig:Re5000ens} the enstrophy during the final parts
of the simulation is given. We note that for the first time the MRT1
system produces less enstrophy (is more dissipative) than the BGK
system with Ehrenfest limiter.

\begin{figure}[h!]
\centering
\includegraphics[width=0.8\textwidth]{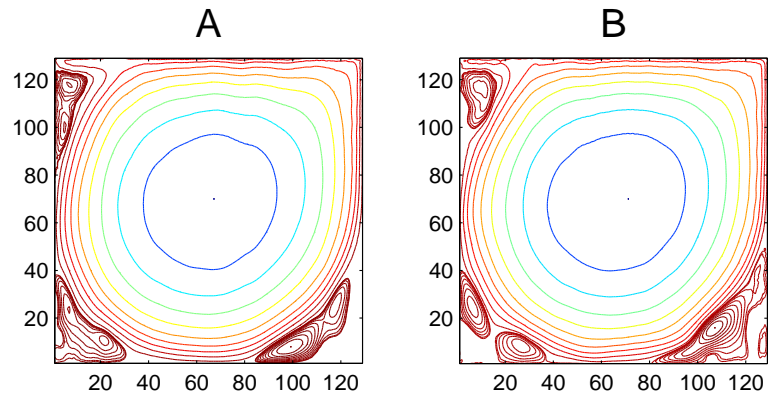}
\caption{Contour plots of stream functions of A: Re7500 and B:
Re10000 BGK + Ehrenfest systems,following 10000000 time steps at
Re5000. \label{fig:remconts}}
\end{figure}

\begin{figure}[h!]
\centering
\includegraphics[width=0.5\textwidth]{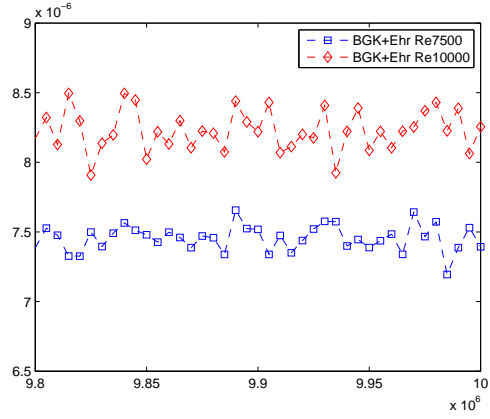}
\caption{Enstrophy in the  Re7500 and  Re10000 BGK + Ehrenfest
systems during the final $2\cdot 10^5$ time steps\label{fig:remens}}
\end{figure}

For the final two Reynolds numbers we use, 7500 and 10000, only the
BGK system with the Ehrenfest limiter completes the simulation. The
corresponding streamfunction plots are given in Figure
\ref{fig:remconts} and they exhibit multiple vortices in the corners
of the domain.

The enstrophy plots are given in Figure \ref{fig:remens}, as the
theoretical Reynolds number increases so does the level of
enstrophy.

\subsection{First Hopf Bifurcation}
As Reynolds number increases the flow in the cavity is no longer
steady and a more complicated flow pattern emerges. On the way to a
fully developed turbulent flow, the lid-driven cavity flow is known
to undergo a series of period doubling Hopf bifurcations.

A survey of available literature reveals that the precise value of
$\mathrm{Re}$ at which the first Hopf bifurcation occurs is somewhat
contentious, with most current studies (all of which are for
incompressible flow) ranging from around
$\mathrm{Re}=7400$--$8500$~\cite{bruneau06,pan00,peng03}. Here, we
do not intend to give a precise value because it is a well observed
grid effect that the critical Reynolds number increases (shifts to
the right) with refinement (see, e.g., Fig.~3 in~\cite{peng03}).
Rather, we will be content to localise the first bifurcation and, in
doing so, demonstrate that limiters are capable of regularising
without effecting fundamental flow features.

To localise the first bifurcation we take the following algorithmic
approach. Entropic equilibria are in use. The initial uniform fluid
density profile is $\rho=1.0$ and the velocity of the lid is
$u_0=1/10$ (in lattice units). We record the unsteady velocity data
at a single control point with coordinates $(L/16,13L/16)$ and run
the simulation for $5000L/u_0$
time steps. Let us denote the final 1\% of this signal by
$(u_\mathrm{sig},v_\mathrm{sig})$. We then compute the \emph{energy}
$E_u$ ($\ell_2$-norm normalised by non-dimensional signal duration)
of the deviation of $u_\mathrm{sig}$ from its mean:
\begin{equation}\label{energy}
    E_u :=   \biggl\| \sqrt{\frac{L}{u_0 |u_{\mathrm{sig}}|}} (u_{\mathrm{sig}}- \overline{u_{\mathrm{sig}}}) \biggr\|_{\ell_2},
\end{equation}
where $|u_{\mathrm{sig}}|$ and $\overline{u_{\mathrm{sig}}}$ denote
the length and mean of $u_{\mathrm{sig}}$, respectively. We choose
this robust statistic instead of attempting to measure signal
amplitude because of numerical noise in the LBM simulation. The
source of noise in LBM is attributed to the existence of an
inherently unavoidable neutral stability direction in the numerical
scheme (see, e.g.,~\cite{Brownlee2}).

We opt not to employ the ``bounce-back'' boundary condition used in
the previous steady state study. Instead we will use the diffusive
Maxwell boundary condition (see, e.g.,~\cite{cercignani75}), which
was first applied to LBM in~\cite{ansumali02}. The essence of the
condition is that populations reaching a boundary are reflected,
proportional to equilibrium, such that mass-balance (in the bulk)
and detail-balance are achieved. The boundary condition coincides
with ``bounce-back'' in each corner of the cavity.

To illustrate, immediately following the advection of populations
consider the situation of a wall, aligned with the lattice, moving
with velocity $u_\mathrm{wall}$ and with outward pointing normal to
the wall in the negative $y$-direction (this is the situation on the
lid of the cavity with $u_\mathrm{wall}=u_0$). The implementation of
the diffusive Maxwell boundary condition at a boundary site $(x,y)$
on this wall consists of the update
\begin{equation*}
    f_i(x,y,t+1) = \gamma f^{*}_i(u_\mathrm{wall}),\qquad i=4,7,8,
\end{equation*}
with
\begin{equation*}
  \gamma = \frac{f_2(x,y,t)+f_5(x,y,t)+f_6(x,y,t)}{f^{*}_4(u_\mathrm{wall})+
  f^{*}_7(u_\mathrm{wall})+f^{*}_8(u_\mathrm{wall})}.
\end{equation*}
Observe that, because density is a linear factor of the
equilibria, the density of the wall is
inconsequential in the boundary condition and can therefore be taken
as unity for convenience. As is usual, only those populations
pointing in to the fluid at a boundary site are updated. Boundary
sites do not undergo the collisional step that the bulk of the sites
are subjected to.

We prefer the diffusive boundary condition over the often preferred
``bounce-back'' boundary condition with constant lid profile. This
is because we have experienced difficulty in separating the
aforementioned numerical noise from the genuine signal at a single
control point using ``bounce-back''. We remark that the diffusive
boundary condition does not prevent unregularised LBGK from failing
at some critical Reynolds number.

\begin{figure}
\begin{centering}
\includegraphics[width=0.5\textwidth]{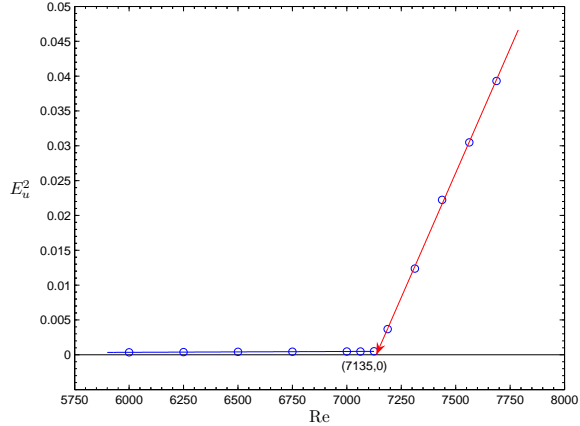}
\caption{Plot of energy squared, $E_u^2$~\eqref{energy}, as a
function of Reynolds number, $\mathrm{Re}$, using LBGK regularised
with the median filter limiter with $\delta = 10^{-3}$ on a $100
\times 100$ grid. Straight lines are lines of best fit. The
intersection of the sloping line with the $x$-axis occurs close to
$\mathrm{Re}=7135$. \label{ERe}}
\end{centering}
\end{figure}

\begin{figure}
\begin{centering}
\includegraphics[width=0.5\textwidth]{{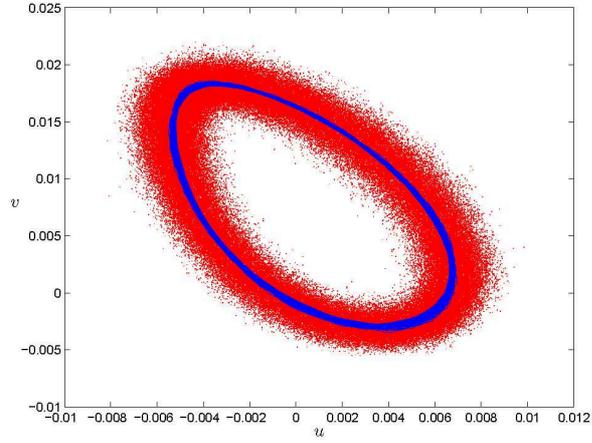}}
\caption{A phase trajectory for velocity components for the signal
$(u_\mathrm{sig},v_\mathrm{sig})$ at the monitoring point
$(L/16,13L/16)$ using LBGK regularised with the median filter
limiter with $\delta = 10^{-3}$ on a $100 \times 100$ grid
($\mathrm{Re}=7375$). Dots represent simulation results at various
time moments and the solid line is a $100$ step time average of the
signal. \label{phase}}
\end{centering}
\end{figure}

\begin{figure}
\begin{centering}
\includegraphics[width=0.5\textwidth]{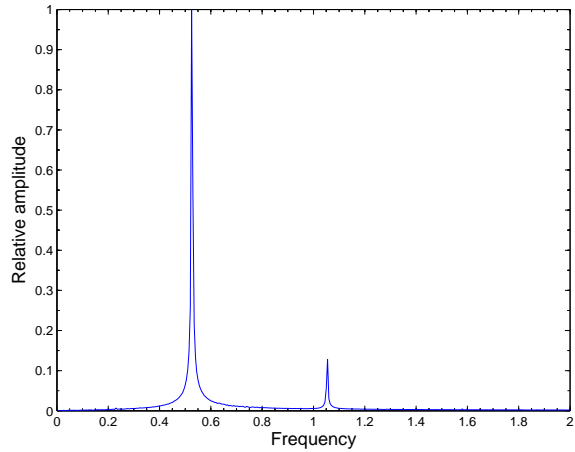}
\caption{Relative amplitude spectrum for the signal $u_\mathrm{sig}$
at the monitoring point $(L/16,13L/16)$ using LBGK regularised with
the median filter limiter with $\delta = 10^{-3}$ on a $100 \times
100$ grid ($\mathrm{Re}=7375$). We measure a dominant frequency of
$\omega=0.525$.\label{spec}}
\end{centering}
\end{figure}

Now, we conduct an experiment and record~\eqref{energy} over a range
of Reynolds numbers. In each case the median filter limiter is
employed with parameter $\delta=10^{-3}$. Since the transition
between steady and periodic flow in the lid-driven cavity is known
to belong to the class of standard Hopf bifurcations we are assured
that $E_u^2\propto \mathrm{Re}$~\cite{ghadder86}. Fitting a line of
best fit to the resulting data localises the first bifurcation in
the lid-driven cavity flow to $\mathrm{Re}=7135$ (Fig.~\ref{ERe}).
This value is within the tolerance of $\mathrm{Re}=7402\pm4\%$ given
in~\cite{peng03} for a $100 \times 100$ grid. We also provide a
(time averaged) phase space trajectory and Fourier spectrum for
$\mathrm{Re}=7375$ at the monitoring point (Fig.~\ref{phase} and
Fig.~\ref{spec}) which clearly indicate that the first bifurcation
has been observed.

\rubric{9. CONCLUSION} \setcounter{section}{9}
\setcounter{subsection}{0} \setcounter{subsubsection}{1}

In the 1D shock tube we do not find any evidence that
maintaining the proper balance of entropy (implementing ELBM)
regularizes spurious oscillations in the LBM. We note that
entropy production controlled by $\alpha$ and viscosity
controlled by $\beta$ are composite in the collision integral
(\ref{eq:CollisionIntegral}). A weak lower approximation to
$\alpha$ would lead  effectively to addition of dissipation at
the mostly far from equilibrium sites and therefore would
locally increase viscosity. Therefore the choice of the method
to implement the entropic involution is crucial in this scheme.
Any method which is not sufficiently accurate could give a
misleading result.

In the 2D lid driven cavity test we observe that implementing
TRT\cite{Ginzburg} or MRT\cite{Lallemand} with certain
relaxation rates can improve stability. The increase in
stability from using TRT can be attributed to the correction of
the numerical slip on the boundary, as well as increasing
dissipation. What is the best set of parameters to choose for
MRT is not a closed question. The parameters used in this work
originally proposed by Lallemand and Luo \cite{Lallemand} are
based on a linear stability analysis. Certain choices of
relaxation parameters may improve stability while qualitatively
changing the flow, so parameter choices should be justified
theoretically, or alternatively the results of simulations
should be somehow validated. Nevertheless the parameters used
in this work exhibit an improvement in stability over the
standard BGK system.

Modifying the relaxation rates of the different modes changes
the production of dissipation of different components at
different orders of the dynamics. The higher order dynamics of
latttice Boltzmann methods include higher order space
derivatives of the distribution functions. MRT could exhibit
the very nice property that where these derivatives are near to
zero that MRT has little effect, while where these derivatives
are large (near shocks and oscillations) that additional
dissipation could be added, regularizing the system.

Using entropic limiters explicitly adds dissipation
locally\cite{Brownlee}. The Ehrenfest steps succeed to
stabilize the system at Reynolds numbers where other tested
methods fail, at the cost of the smoothness of the flow. We
also implemented an entropic limiter using MRT technology. This
also succeeded in stabilizing the system to a degree, however
the amount of dissipation added is less than an Ehrenfest step
and hence it is less effective. The particular advantage of a
limiter of this type over the Ehrenfest step is that it can
preserve the correct production of dissipation on physical
modes across the system. Other MRT type limiters can easily be
invented by simply varying the relaxation parameters.

As previously mentioned there have been more filtering
operations proposed\cite{Ricot}. These have a similar idea of
local (but not pointwise) filtering of lattice Boltzmann
simulations. A greater variety of variables to filter have been
examined, for example the macroscopic field can be filtered
rather than the mesoscopic population functions.

We can use the Enstrophy statistic to measure effective
dissipation in the system. The results from the lid-driven
cavity experiment indicate that increased total dissipation
does not necessarily increase stability. The increase in
dissipation needs to be targeted onto specific parts of the
domain or specific modes of the dynamics to be effective.

Using the global median filter in the lid-driven cavity we find
that the expected Reynolds number of the first Hopf bifurcation
seems to be preserved, despite the additional dissipation. This
is an extremely positive result as it indicates that if the
addition of dissipation needed to stabilize the system is added
in an appropriate manner then qualitative features of the flow
can be preserved.

Finally we should note that the stability of lattice Boltzmann
systems depends on more than one parameter. In all these
numerical tests the Reynolds number was modified by altering
the rate of production of shear viscosity. In particular, for
the lid driven cavity the Reynolds number could be varied by
altering the lid speed, which was fixed at 0.1 in all of these
simulations. Since the different modes of the dynamics include
varying powers of velocity, this would affect the stability of
the system in a different manner to simply changing the shear
viscosity coefficient. In such systems the relative
improvements offered by these methods over the BGK system could
be different.

The various LBMs all work well in regimes where the macroscopic
fields are smooth. Each method has its limitation as they
attempt to simulate flows with, for example, shocks and
turbulence, and it is clear that something needs to be done in
order to simulate beyond these limitations. We have examined
various add-ons to well-known implementations of the LBM, and
explored their efficiency. We demonstrate that add-ons based on
the gentle modification of dissipation can significantly expand
the stable boundary of operation of the LBM.

\vspace{.4cm}
\rubric{REFERENCES}
%
%
%
\vspace{-1.5cm}
\renewcommand\refname{}
\bibliographystyle{alpha}

\end{document}